\documentclass[twocolumn]{aastex62}
\usepackage{hyperref,astrojournals,amsmath,amssymb,mathtools,graphicx,txfonts,microtype,nicefrac,xspace}
\usepackage[switch]{lineno}
\usepackage{enumitem,kantlipsum}

\renewcommand*{\vec}[1]{\ensuremath{\boldsymbol{#1}}}
\let\oldhat\hat
\renewcommand*{\hat}[1]{\oldhat{\vec{#1}}}
\newcommand*{\der}[2]{\ensuremath{\frac{d #1}{d #2}}\xspace}
\newcommand*{\subtwo}[2]{\ensuremath{{#1}_{\text{#2}}\xspace}}
\newcommand*{\sub}[3]{\ensuremath{{#1}_{\text{#2}}^{\text{(#3)}}\xspace}}

\newcommand{\unitvectorslope}[1]{\ensuremath{%
\frac{\c{\oldhat{#1}}{z}}%
     {\sqrt{1-\c{\oldhat{#1}}{z}^2}}}}
\renewcommand{\c}[2]{\ensuremath{#1_{\text{#2}}}} 	
\newcommand{\ia}{\c{i}{a}\xspace}
\newcommand{\ib}{\c{i}{b}\xspace}
\newcommand{\ie}{\c{i}{e}\xspace}
\newcommand{\Md}{M_{\text{disk}}}

\newcommand{\Gyr}{{\,\rm Gyr}}
\newcommand{\Myr}{{\,\rm Myr}}

\renewcommand*{\vec}[1]{\boldsymbol{#1}}

\hypersetup{%
  pdftitle={Disks},
  pdfauthor={\textcopyright\ authors},
  bookmarksopen=true,
  colorlinks=true,
  linkcolor=black,
  citecolor=black,
  urlcolor=black}

\begin{document}

\title{On the Dynamics of the Inclination Instability}
\shorttitle{Inclination Instability} 
\shortauthors{Madigan, Zderic, McCourt \& Fleisig}
 
\author{Ann-Marie Madigan}
\affiliation{JILA and Department of Astrophysical and Planetary Sciences, CU Boulder, Boulder, CO 80309, USA}
\email{annmarie.madigan@colorado.edu}

\author{Alexander Zderic}
\affiliation{JILA and Department of Astrophysical and Planetary Sciences, CU Boulder, Boulder, CO 80309, USA}

\author{Michael McCourt}
\affiliation{Astronomy Department, Broida Hall, University of California, Santa Barbara, CA 93106, USA}

\author{Jacob Fleisig}
\affiliation{JILA and Department of Astrophysical and Planetary Sciences, CU Boulder, Boulder, CO 80309, USA}

\begin{abstract}
Axisymmetric disks of eccentric Kepler orbits are vulnerable to an instability that causes orbits to exponentially grow in inclination, decrease in eccentricity, and cluster in their angle of pericenter. Geometrically, the disk expands to a cone shape that is asymmetric about the mid-plane. In this paper, we describe how secular gravitational torques between individual orbits drive this ``inclination instability". We derive growth timescales for a simple two-orbit model using a Gauss $N$-ring code, and generalize our result to larger $N$ systems with $N$-body simulations. We find that two-body relaxation slows the growth of the instability at low $N$ and that angular phase coverage of orbits in the disk is important at higher $N$. As $N \to \infty$, the e-folding timescale converges to that expected from secular theory. \\

\end{abstract}


\section{Introduction}
\label{sec:intro}
In \citet{Madigan2016}, we introduced a new dynamical instability in axisymmetric Kepler disks of eccentric orbits.  We showed that secular (orbit-averaged) gravitational torques between orbits in the disk drive exponential growth of their inclinations.  As the orbits' inclinations grow, they tilt in the same way with respect to the disk plane. This leads to clustering in their angles of pericenter and the initially thin disk expands into a cone shape. Concurrently, the orbital eccentricities decrease and perihelion distances increase. We proposed that this could be at work in the outer solar system between minor planets scattered to large orbital eccentricities by the giant planets. It can explain the observed clustering in the angle of pericenter and detached, high perihelia objects \citep{Trujillo2014}. 

This paper serves to explain the physics behind the inclination instability. 
In section~\ref{sec:theory} we describe the mechanism of the instability through a simple, yet predictive, two-orbit toy model. We show analytically that inclinations grow exponentially in disks of eccentric orbits, and numerically derive the associated growth timescale. We discuss our results and their application to the outer solar system in section~\ref{sec:discussion} .

\section{Mechanism of Instability}
\label{sec:theory}

We consider a thin, axisymmetric disk of particles on eccentric orbits ($e \gtrsim 0.6$) around a much more massive central object.  All orbits take part in the inclination instability which changes their orientation.  However, to illustrate the instability's physical mechanism, we consider just two orbits within the disk and see how this system responds to a small perturbing force. The response of the system to the small perturbing force is best described with a set of angles different than the traditionally used Kepler elements (inclination $i$, longitude of ascending node $\Omega$, and argument of pericenter $\omega$). These new angles are \ia, \ib, and \ie, and we originally introduced them in  \citet{Madigan2016}.  The angles represent rotations of the orbit about its semi-major (${\hat a}$) axis, semi-minor (${\hat b} \equiv \hat{j} \times \hat{a}$) axis, and angular momentum vector ($\hat{j}$), respectively,\footnote{Note that these are not true rotation angles. See the Appendix for more information.}, i.e.,
\begin{subequations}
\begin{align}
  \ia &= \arctan\left[\unitvectorslope{b}\right], \\
  \ib &= \arctan\left[-\unitvectorslope{a}\right], \\
  \ie &= \arctan\left[\oldhat{a}_{\text{y}}, \oldhat{a}_{\text{x}}\right].
\end{align}
\label{eq:iaibie}
\end{subequations}
\\
The subscripts $x$, $y$, and $z$ denote an inertial Cartesian reference frame with unit vectors, $\hat{x}$, $\hat{y}$, and $\hat{z}$.
The angles \ia, \ib, and \ie are equivalent to that of an aircraft's roll, pitch and yaw. 
We give a graphical representation of \ia and \ib in figure~\ref{fig:twoOrbitToyModel} and convert between these variables and Kepler elements in the Appendix. 

\subsection{Two-orbit toy model}
\label{subsec:toymodel}

\begin{figure}
  \includegraphics[width=\linewidth]{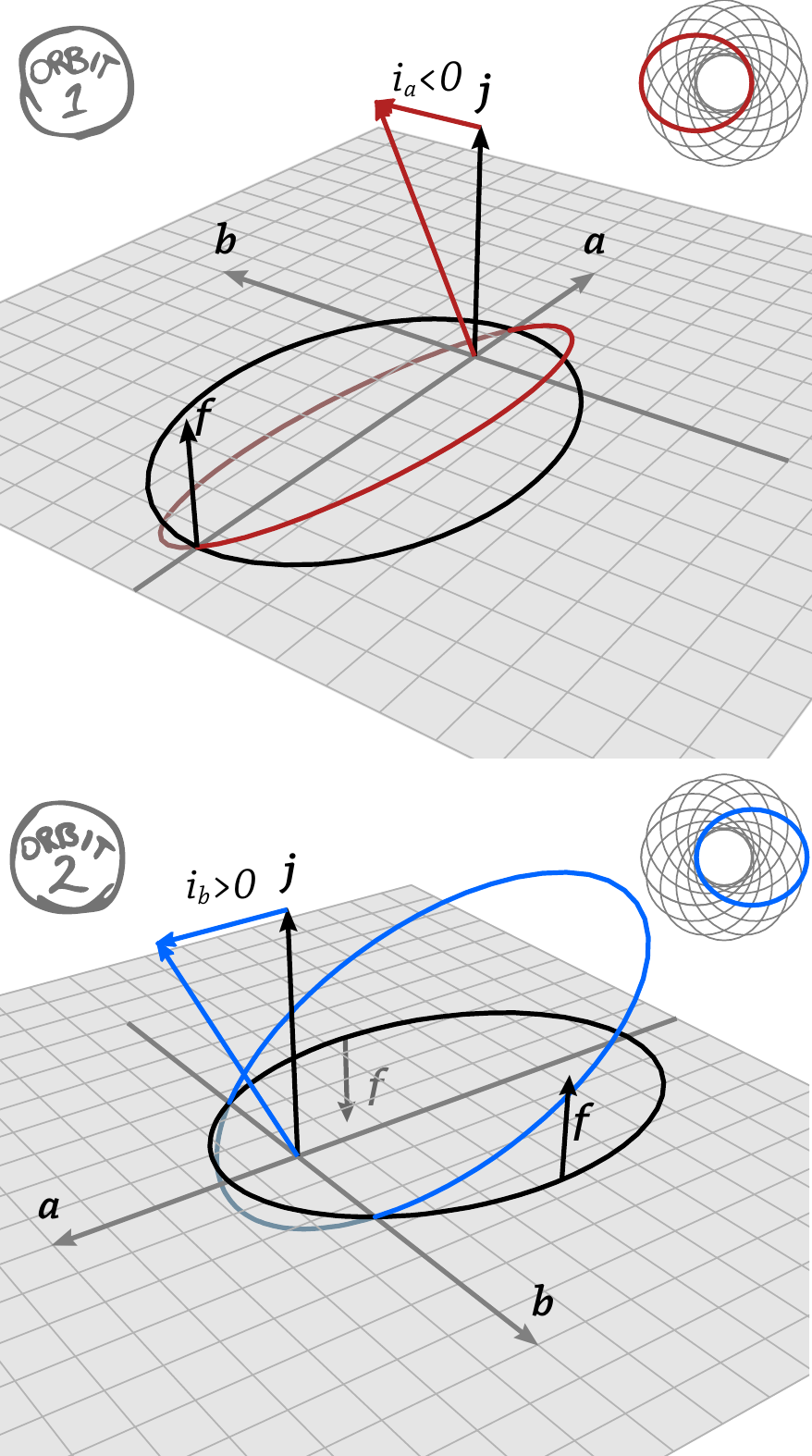}
  \caption{{Idealized, ``two-orbit'' toy model for the inclination
      instability.  The top right of each panel shows the location of the orbit in the disk from a face-on perspective.  
      (\textit{top}): Orbit~1 experiences a net upward
      force $\vec{f}$.  This force produces a torque along the $\hat{b}$
      axis, rotating the orbital plane such that $\ia<0$.
      (\textit{bottom}): Diametrically opposed orbit~2 feels a force
      due to the rotation of orbit~1; this force produces a torque along the
      $\hat{a}$ direction of orbit~2, rotating its orbital plane such that
      $\ib>0$.}}\label{fig:twoOrbitToyModel}
\end{figure}
\begin{figure}
\begin{center}
   \includegraphics[width=\columnwidth]{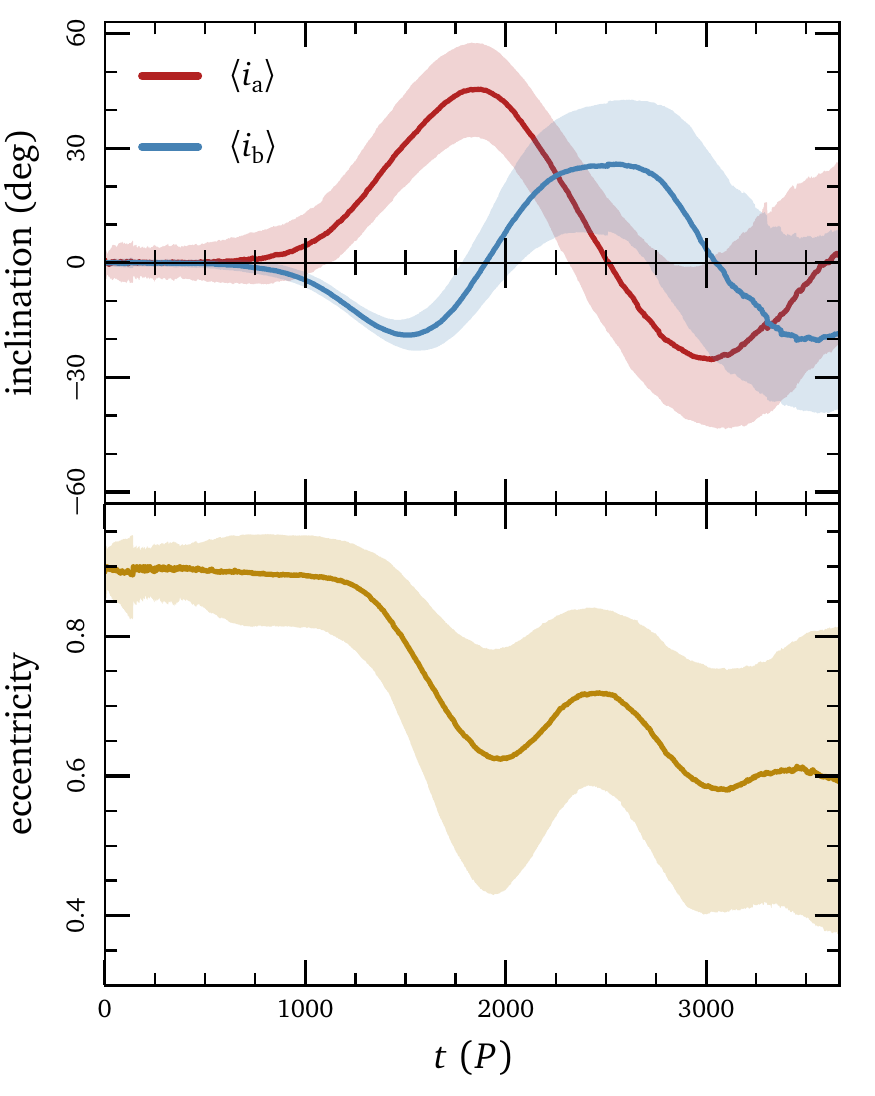}
\caption{(\textit{top}): Median values of angles \ia and \ib for all particles in a disk as a function of time in units of orbital period. Shaded regions indicate one-sigma quantile contours.
	Both \ia and \ib rise exponentially at early times. As predicted by the two-orbit toy model, \ia and \ib increase with opposite sign.
	 (\textit{bottom}): Median value of orbital eccentricity for the same particles. As inclinations increase, eccentricities decrease.}
\label{fig:iaib}
\end{center}
\end{figure}
We consider how two initially coplanar, anti-aligned eccentric orbits respond to a perturbing force, as depicted in figure~\ref{fig:twoOrbitToyModel}.

\begin{enumerate}
\item We perturb orbit~1 with a uniform force in the
  $+\hat{z}$-direction. Such a force could arise from, e.\,g., a
  temporary fluctuation in the center of mass of the disk.
  Because of the orbit's symmetry around the ${\hat a}$ axis, all torques in the ${\hat a}$-direction cancel.
  However, because of the orbit's asymmetry around the ${\hat b}$ axis, the vertical
  force perturbation creates a net torque in the ${\hat b}$-direction.
  This asymmetry between the torques is further exacerbated by the time-averaged torque, $\overline{\vec{\tau}} = \overline{\vec{r} \times \vec{f}}$, at apocenter
  being greater than at pericenter because the particle
  spends a greater fraction of its time at apocenter, ($\sim[(1+e)/(1-e)]^{3/2}$).
  Thus, the net time-averaged torque points in the
  $+\hat{b}$-direction.  Consequently, the angular momentum vector
  $\vec{j}$ of orbit~1 rotates counter clockwise about its semi-major
  axis, rolling the orbital plane such that $\ia{}<0$.

\item The roll of orbit~1 induces out-of-plane forces along the
  sides of orbit~2.  These forces are asymmetric around the ${\hat a}$ axis and 
  imply a time-averaged torque on orbit~2 in its $+\hat{a}$-direction.  This torque rotates
  orbit~2 about its semi-minor axis, pitching the orbital plane such
  that $\ib{}>0$.

\item This rotation raises orbit~2's apocenter, and thus its center of
  mass, above the original orbital plane.  The raising of orbit~2's apocenter 
  raises the center of mass of the disk, which reinforces the initial 
  perturbation. Thus, the disk is unstable to the initial perturbation, and 
  the inclination will run away to large values.
  
  \end{enumerate}

As the orbits incline off the disk plane, conservation of angular momentum requires that they decrease in eccentricity. They do so via the same secular gravitational torques that drive growth in inclination. Let us take orbit~1 as an example. After orbit~1 begins to roll over its major axis, its angular momentum vector is no longer parallel to the initial perturbing force. A component of the force acts therefore in its orbital plane. This component causes a positive time-averaged torque resulting in a decrease in eccentricity. A similar analysis shows that orbit~2 must also decrease in eccentricity. \\

Though we have only considered two orbits in the disk, all orbits take part in the instability. We show this in figure~\ref{fig:iaib} in which we plot the median values of \ia and \ib (top), and eccentricity (bottom) for a representative $N$-body simulation. We refer the reader to figure~\ref{fig:timescale-n} to clearly see exponential growth of \ib. Here we use the  {\tt REBOUND} code \citep{Rein2012}, with the IAS15 integrator \citep{Rein2015}.   
We distribute $N=100$ massive particles in a thin axisymmetric disk in the potential of a central body of mass $M$. The orbit eccentricities are initialized to $e = 0.9$, semi-major axes are uniformly distributed in the narrow range [$0.9, 1.1$].  Mean anomaly, argument of pericenter, and angle of ascending node are distributed uniformly in ($0, 2\pi$]. 

The two-orbit toy model makes several testable predictions confirmed by $N$-body experiments (see also figure 2 of \citet{Madigan2016}):
\begin{enumerate}
\item If the magnitudes of the torques are linear in \ia and \ib, as one might expect for small deviations from an equilibrium, this instability should grow exponentially in time.
\item The orbits should grow their $|\ia|$ and $|\ib|$ values with opposite signs and \ib/\ia in constant ratio. 
\item A constant ratio \ib/\ia implies a constant angle of pericenter, as for small inclinations $\omega(\ia,\ib) \sim \arctan{ |\ib/\ia|}$ ($+\pi$ if \ia $< 0$). As \ia and \ib depend on the eccentricities of orbits, $\omega(\ia,\ib)$ should cluster to one of two eccentricity-dependent values, depending on the direction of the initial perturbing force. 
\end{enumerate}

\subsection{Growth timescale of instability}
\label{subsec:gauss-code}
We estimate the e-folding timescale of the instability between two orbits as follows.
For a given orbit with its $\vec{\hat{a}}$ and $\vec{\hat{b}}$ axes, two components of the torque that rotate its orbital plane are:
\begin{subequations}
\begin{align}
  \subtwo{\tau}{a} &\equiv \vec{\tau}\cdot\vec{\hat{a}}, \\
  \subtwo{\tau}{b} &\equiv \vec{\tau}\cdot\vec{\hat{b}}.
\end{align}
\end{subequations}
We denote the torque on orbit~2 due to orbit~1 as \sub{\tau}{}{2,1} and the torque on orbit~1 due to orbit~2 as \sub{\tau}{}{1,2}.
Torque acting on orbit 2 can be expressed  as
\begin{equation}\label{eq:vec-rot}
\sub{\tau}{a}{2,1} (\sub{i}{a}{1}) = \der{\sub{j}{}{2} (\sub{i}{b}{2}) }{t} = \sub{j}{}{2}  \der{\sub{i}{b}{2} }{t}, 
\end{equation}
(where the last step is simply vector rotation), and expanded via Taylor series and linearized 
\begin{equation}\label{eq:TS}
\sub{\tau}{a}{2,1} (\sub{i}{a}{1}) = \left(\der{\sub{\tau}{a}{2,1}}{\sub{i}{a}{1}}\right) \sub{i}{a}{1}. 
\end{equation}
Since \ia represents a rotation of angular momentum in the ($-\vec{\hat{b}}$) direction and \subtwo{i}{b} represents a rotation of angular momentum in the $\vec{\hat{a}}$ direction, we have the linearized equations of motion:
\begin{subequations}\label{eq:linear-eom}
\begin{align}
  \sub{j}{}{2} \der{\sub{i}{b}{2}}{t} &=  \sub{\tau}{a}{2,1} (\sub{i}{a}{1})
  = \left(\der{\sub{\tau}{a}{2,1}}{\sub{i}{a}{1}}\right) \sub{i}{a}{1}
  = \alpha\,\sub{j}{}{2}\,\sub{i}{a}{1}, \quad\textrm{and} \label{eq:linear-eom-a} \\
  \sub{j}{}{1} \der{\sub{i}{a}{1}}{t} &= -\sub{\tau}{b}{1,2} (\sub{i}{b}{2})
  = -\left(\der{\sub{\tau}{b}{1,2}}{\sub{i}{b}{2}}\right) \sub{i}{b}{2}
  = -\beta\,\sub{j}{}{1}\,\sub{i}{b}{2}, \label{eq:linear-eom-b}
\end{align}
\end{subequations}
where in the last step we have defined the coefficients
\begin{subequations}\label{eq:alpha-beta-defs}
\begin{align}
  \alpha &\equiv \frac{1}{\sub{j}{}{2}} \left(\der{\sub{\tau}{a}{2,1}}{\sub{i}{a}{1}}\right)
  = \frac{1}{\sub{j}{}{1}} \left(\der{\sub{\tau}{a}{1,2}}{\sub{i}{a}{2}}\right), \label{eq:alpha-def} \\
  \beta &\equiv \frac{1}{\sub{j}{}{1}} \left(\der{\sub{\tau}{b}{1,2}}{\sub{i}{b}{2}}\right)
  = \frac{1}{\sub{j}{}{2}} \left(\der{\sub{\tau}{b}{2,1}}{\sub{i}{b}{1}}\right). \label{eq:beta-def}
  \end{align}
\end{subequations}
There are similar expressions for $\sub{i}{a}{1}{}'(t)$ and $\sub{i}{b}{2}{}'(t)$. In the following, we assume the parameters $\alpha$ and $\beta$ are constants, which is consistent with $N$-body simulations of the instability. 
We then differentiate equation~\ref{eq:linear-eom-a} and combine it with equation~\ref{eq:linear-eom-b} to yield an equation of motion for \ib:
\begin{align}
  \sub{i}{b}{2}{}''(t) = \alpha\,\sub{i}{a}{1}{}'(t) = - \alpha\beta\,\sub{i}{b}{2}.
  \label{eq:eom}
\end{align}
For small inclinations, the inclination angles evolve in time as
\begin{align}
  \{\ia,\ib\} &\propto \exp(\gamma t)\label{eq:ib-exp-gamma},\\
  \gamma &\equiv \sqrt{-\alpha\beta}\label{eq:growth-rate},
\end{align}
where $\gamma$ is the growth rate of the instability. 
Equation~\ref{eq:eom} shows that the instability exponentially grows orbit inclinations through mutual secular gravitational torques between eccentric orbits. 

We numerically evaluate $\alpha$ and $\beta$ from equation \ref{eq:alpha-beta-defs} using Gauss' approximation, as described in \citet{Gur07}.\footnote{Our source code is publicly available at \url{https://github.com/mkmcc/orbit-torque}.}
We estimate the time-averaged torques between two orbits by smearing the bodies out into ``wires'' with a linear mass density along the orbit inversely proportional to the instantaneous Kepler velocity; we then split each wire into $n$ segments equally spaced in eccentric anomaly, and we directly sum the $n^2$ gravitational torques of all pairs of segments.
We repeatedly double the number of segments until the total torque converges.
Though our numeric estimate for the time-averaged torques between orbits does not require gravitational softening, we include it in our calculation as arbitrarily large torques can arise for certain orientations of the pair of orbits.
This is unsatisfactory from a theoretical perspective because the large torques arising over small spatial scales destroy the linearity presumed in equation~\ref{eq:linear-eom}. 
These large torques are also unphysical, because they would drive scattering or diffusion of the orbits on a timescale shorter than the instability growth timescale.
They therefore should not be included in our calculation, and gravitational softening is one way to approximate this behavior.
We adopt a fiducial force-softening length $s=(3\times10^{-2})\,a$, chosen to exclude close encounters that would drive diffusion on a timescale shorter than $1/\gamma$. With this choice, the linear theory we present applies over the majority of the parameter space. The  boundaries between stable and unstable regions do not depend significantly on our chosen value of $s$. However, the maximum growth rate scales with the softening length as $\gamma_{\text{max}} \sim s^{-3/4}$.

\begin{figure}
  \centering     \includegraphics[width=\columnwidth]{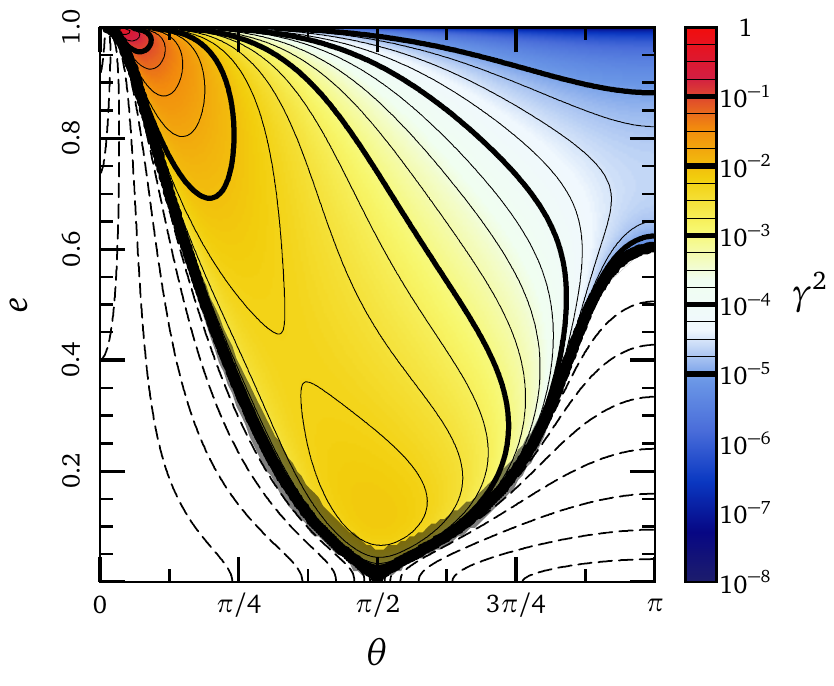}
  \caption{Squared growth rate $\gamma^2$ for two orbits gravitationally torquing each other. This is estimated from the linearized equations of motion for the system (equation~\ref{eq:growth-rate}), as a function of orbital eccentricity and angular separation between orbits in the plane, $\theta$. 
Only the unstable region with $\gamma^2>0$ is colored. 
The white, uncolored region of the plot with dashed contours is stable. The thick, gray band near the $\gamma^2=0$ boundary marks the region in which the linear approximation in equation~\ref{eq:linear-eom} breaks down, causing an order-unity uncertainty in our estimate of the timescale.
This region also depends on the softening length we adopt.}
  \label{fig:growth-rate}
\end{figure}

Using this numerical approach, we calculate the squared growth rate, $\gamma^2$, as a function of orbital eccentricity and angular separation, $\theta = \arccos ( \vec{\c{\hat e}{1}} \cdot \vec{\c{\hat e}{2}} )$. We initialize the equal-mass orbits with similar semi-major axes. The growth rate is normalized to the orbital period, so that $\gamma$ is the number of e-foldings per orbit. We show $\gamma^2$ as a function of $\theta$ and $e$ in figure~\ref{fig:growth-rate}. 
At every eccentricity, the orbits are unstable in some range of angular separations. The largest growth rates occur at the highest orbital eccentricities and small-to-intermediate angular separations. In general, as eccentricity increases, the range of stable angular configurations shrinks. At very small angular separations, the two orbits are unstable only for very high eccentricities. Interestingly, our results show that the angular separation ($\theta = \pi$) used by the toy model in section~\ref{subsec:toymodel} is the least-unstable unstable region of figure~\ref{fig:growth-rate}. The inclination instability only occurs in high eccentricity disks ($e \gtrsim 0.6$) because below this boundary most orbits are stable with respect to one another; tests with $N$-body simulations confirm this.

There is a marked boundary in $e-\theta$ space between stable and unstable configurations. The location of this boundary can be understood from the two-orbit model: For low $\theta$, orbits exert stabilizing vertical forces on each other because their sides overlap. As $\theta$ increases, the distance between their sides grows until the force acting on each side reverses and becomes unstable. This reversal happens at smaller separations for high eccentricity orbits as the width of an orbit scales as $\sqrt{1 - e^2}$. Orbits of nonzero eccentricity are unstable at $\theta = \pi/2$. As $\theta$ increases towards $\pi$, the growth rate of the instability decreases due to the increasing distance and decreasing gravitational force between the orbits. Low-eccentricity orbits are the most stable as forces at pericenter become stronger than those at apocenter. The squared growth rate $\gamma^2$ is negative in the stable regions which implies that the two orbits may be undergoing oscillatory behavior. We will explore the dynamical behavior of the orbits in the regions of stability in a future study.

\subsection{Generalization to $N$ bodies}
\label{subsec:n-orbits}
To apply the instability to real astrophysical systems with more than two bodies, we need to know the growth rate for large $N$ systems. 
We start with the hypothesis that the inclination instability acts on a secular timescale, 
\begin{align}
	t_{\rm sec} \sim \frac{1} {2\pi} \frac{M}{\Md} P,
	\label{eq:sec-time}
\end{align}
where $P$ is the orbital period. Here $t_{\rm sec}$ is derived as the time it takes to change an orbit's angular momentum by order of its circular angular momentum using a specific torque over one orbital period of $\tau \sim Gm/a$. To test this hypothesis, we use numerical $N$-body simulations with the {\tt REBOUND} code. 
We simulate the instability for a range of disk masses ($\Md = [10^{-2}, 10^{-3}, 10^{-4}]$, where the central mass $M = 1$), and number of particles ($N$ = [100, 200, 500, 1000]), averaging many tens of simulations for each parameter set to reduce the noise. 
Each particle has $m=\Md/N$, $e=0.7$, and $i=10^{-4}$~rad. Semi-major axes are distributed uniformly in [$0.9, 1.1$]. Mean anomaly, argument of pericenter, and angle of ascending node are randomly distributed in ($0,2\pi$]. We do not use gravitational softening. 

\begin{figure}
\begin{center}
   \includegraphics[ angle=0,width=\columnwidth]{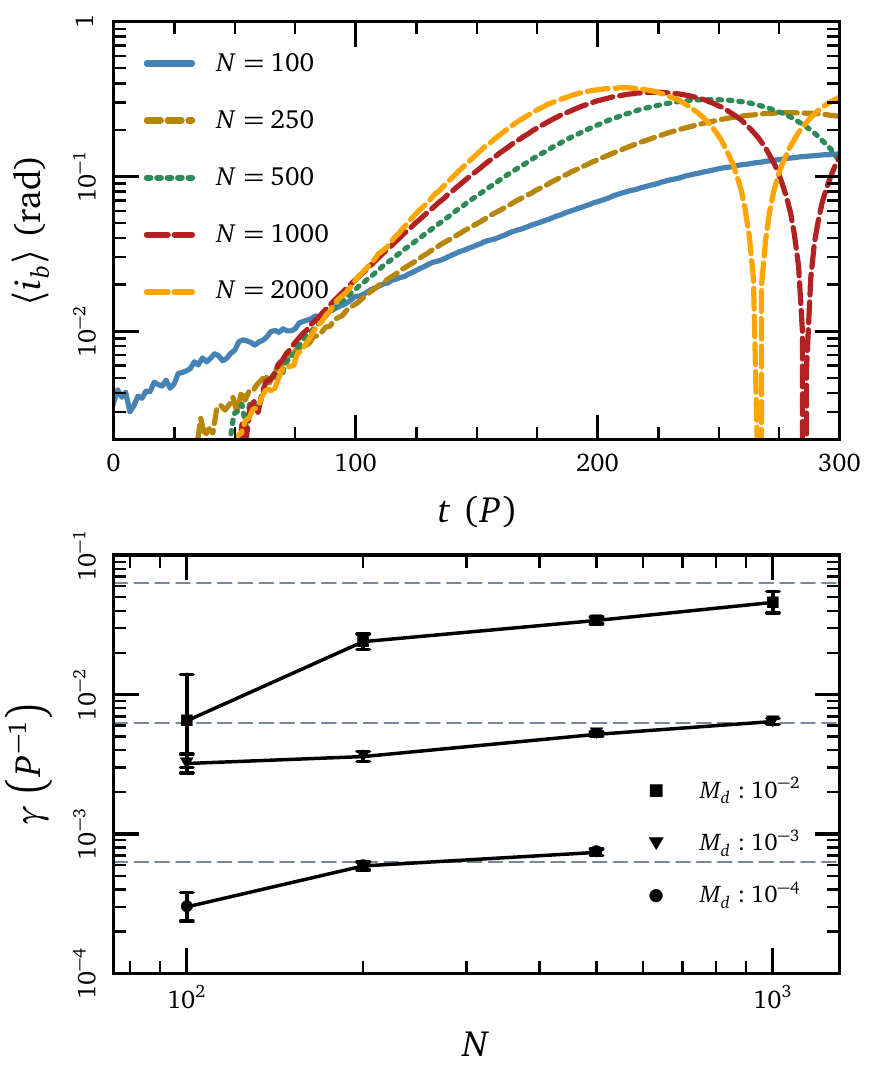}
\caption{(\textit{top}): Exponential growth of the inclination angle \ib for different numbers of disk particles (N) and fixed total mass $\Md = 10^{-2} M$. Time is in units of orbital period at semimajor axis $a = 1$. Note that the slopes (growth rate) of the lines show a strong dependence on $N$. (\textit{bottom}): Instability growth rate ($\gamma$) for different values of $N$ and $\Md$.  The expected secular growth rate ($t_{\rm sec}^{-1}$) is plotted as dashed gray lines (corresponding from top to bottom to disk masses of $\Md = [10^{-2}, 10^{-3}, 10^{-4}]$). Note that the growth rate's dependence on $N$ weakens as $\Md$ is reduced, and ultimately converges to the secular growth rate.}
\label{fig:timescale-n}
\end{center}
\end{figure}

We show the results of our simulations in figure~\ref{fig:timescale-n}. The top plot shows the mean inclination of particles as a function of time for simulations with a fixed total mass $\Md = 10^{-2} M$. The various lines correspond to simulations with differing numbers of particles. A surprising result is how dependent the inclination growth rate is on the number of particles per simulation. The instability proceeds faster with increasing $N$. 
The bottom plot extends this result for different disk masses, showing the growth rate $\gamma$ as a function of $N$ for $\Md = [10^{-2}, 10^{-3}, 10^{-4}]$.  We derive the growth rate in each simulation from the linear parameter of a quadratic model fit to the median $\log\ib$ value\footnote{We use \ib instead of \ia to calculate the exponential growth rate as it is the less noisy parameter (it requires more force for an orbit to change its \ib value as it must lift its apocenter).}.
We optimize the fit parameters with a linear least-squares method and find the error using a bootstrapping method. Our data is Gaussian distributed so a least-squares optimization is equivalent to a maximum likelihood estimator (MLE) approach.

If the e-folding time of the instability matched the secular timescale given by equation \ref{eq:sec-time}, the data in figure~\ref{fig:timescale-n} would align with the dashed gray lines. 
This is clearly not the case however, particularly at low $N$ and large $\Md$. 
We identify three reasons for the deviation of the instability growth rate from a secular one. 
\begin{enumerate}[wide, labelwidth=!, labelindent=0pt]
\item Angular phase coverage of orbits in disk \\
In the two-orbit model, the instability growth rate strongly depends on the angular separation of orbits (see figure~\ref{fig:growth-rate}). In fact, large sections of phase space are stable ($\gamma^2 < 0$). Generalizing to $N$ bodies, the angular phase coverage of orbits in a disk greatly impacts the overall dynamics, including whether or not the disk is stable or unstable. 
Non-axisymmetric disks are stable; a good example being the eccentric nuclear disk of stars in the Andromeda nucleus \citep{Madigan2018}.  
Furthermore, in the limit of small $N$, few angular separations are realized and axisymmetric disks can also be stable; tests show that at $N \lesssim 20$ the instability doesn't occur within the predicted timescale.
As we randomly select the angular orientation of the orbits in our simulations, the unstable region of phase space gets filled at a rate $\propto N^{\nicefrac{1}{2}}$. Assuming the most unstable angular separation dominates the growth rate, we would then expect the growth rate to follow an $N^{\nicefrac{1}{2}}$ dependence. 
\item Apsidal precession \\
In limit of $\Md \to M$, the Kepler orbit approximation breaks down. Orbits precess rapidly due to the gravitational potential of the disk, and  we expect orbit-averaged torques to weaken significantly. Hence the timescale for the instability should strongly deviate from the secular timescale at large $\Md/M$ values, as seen in figure \ref{fig:timescale-n}. 
\item Two-body relaxation \\
In a near-Keplerian system, the two-body relaxation time is given by 
\begin{align}
	t_{\rm rel} \equiv \frac{M^2}{m^2N\ln(\Lambda)}
	\label{eq:two-body}
\end{align}
\citep{Rauch1996}, where $\ln(\Lambda)$ is the Coulomb logarithm. 
Therefore, the ratio of the two-body timescale to the secular timescale is 
\begin{align}
	\frac{t_{\rm rel}}{t_{\rm sec}} \sim \frac{M}{m\ln(\Lambda)} \propto m^{-1}.
	\label{eq:tb-sec-ratio}
\end{align}
As $m$ decreases, the two-body relaxation timescale increases relative to the secular timescale. Conversely, for larger values of $m$, strong two-body scattering perturbs particles off their orbits and slows the rate of the secular instability.  Thus, we expect the growth rate to have a $m^{-1}$ dependence from two-body relaxation. For a fixed $\Md$, this will emerge as an $N^{-1}$ dependence.
\end{enumerate}

All three of these physical effects coexist simultaneously in our simulations. To explore their relative importance, we parameterize deviation from the secular growth rate with an arbitrary function $\alpha$,
\begin{align}
	\gamma^{-1} = \alpha(N, \Md) ~t_{\rm sec}. 
	\label{eq:gr-time}
\end{align}
We fit our simulation data using two different models, $\alpha = aN^{-\nicefrac{1}{2}}+ b$ and $\alpha = aN^{-1}+ b$, for disk phase-coverage and two-body relaxation, respectively. 
The phase coverage model most accurately fits the large $N$ evolution of the instability timescale ($N=200\textrm{--}1000$). Additionally, it predicts that $\alpha$ converges to the same value for all $\Md$ ($b \approx 0.4$).   
However, the two-body scattering model more accurately fits the low $N$ data points. Therefore, we see a mixture of phase coverage and two-body scattering for the $N$ we simulated. 
In both models, the value of the $a$ parameter decreases with $\Md$. That is, the growth rate's dependence on $N$ weakens as $\Md$ decreases. Apsidal precession is the likely source of this additional $\Md$ dependence.
In summary, our exploration into the growth rate of the instability for large $N$, low $\Md$ systems finds that the e-folding timescale is
	\begin{align}
	t_{\rm e-fold} \sim \frac{0.2} {\pi} \frac{M}{\Md} P.
	\label{eq:e-fold-time}
\end{align}

\section{Discussion}
\label{sec:discussion}
In this paper, we describe the mechanism behind the inclination instability in axisymmetric, eccentric Kepler disks as introduced by \citet{Madigan2016}. 
The instability results in an exponential growth of orbital inclinations, a decrease in orbital eccentricities, and clustering in the argument of pericenter during its linear phase. 
We explain how the instability works with a simple toy model, examining the secular torques between just two orbits (section \ref{subsec:toymodel}). Using linearized equations of motion, we analytically demonstrate the exponential growth of the instability and numerically derive the e-folding rate (section \ref{subsec:gauss-code}). We expand to large $N$ systems (section \ref{subsec:n-orbits}) with a suite of $N$-body simulations.  Our results show that two-body scattering, angular phase coverage of orbits within the disk, and apsidal precession reduce the growth rate of the instability. At low $N$ ($ \approx {10^2}$), two-body scattering is the dominant physical process decreasing the growth rate. At large $N$, the angular phase coverage of orbits is important.  Apsidal precession further decreases the growth rate as $\Md \to M$. 
	This result is general to all $N$-body simulations involving secular dynamical behavior. Thus, we conclude that $N$-body simulations (which for computational efficiency typically use low $N$ and high $\Md/M$ values) generally underestimate the strength of secular dynamics. 

\subsection{Outer Solar System}
\label{subsec:oSS}

The results of section \ref{subsec:n-orbits} can be used to apply the inclination instability to real astrophysical systems that are currently impossible to directly simulate. 
Studies of inner Oort cloud formation in a stellar cluster suggest that 1\textrm{-}10 Earth masses of cometary material may exist at hundreds of au \citep{Brasser2012}. The comets are scattered outward by Jupiter and Saturn, and their pericenters are later decoupled from the planets by perturbations from cluster gas and nearby stars. 
We compare important dynamical timescales for bodies in this region, taking the orbital parameters of the minor planet Sedna as an example \citep[][$e \approx 0.85$, $a \approx 507$ au, $i \approx 12^\circ$]{Brown04}. 

The induced motion of Sedna's orbit due to the four giant planets can be approximately calculated by assigning to the Sun an artificial quadrupole moment, 
\begin{align}
	J_2 = \frac{1}{2} \sum_i^4  \frac{m_i}{M_\odot} \left(\frac{a_i}{R_\odot}\right)^2. 
	\label{eq:j2}
\end{align}
The apsidal and nodal precession frequencies are then given by
\begin{align}
	\dot\omega = \frac{3 J_2}{4} n \left(\frac{R_\odot}{a}\right)^2 \left(\frac{5 \cos^2 i - 1}{(1 - e^2)^2}\right), 
	\label{eq:j2-apse}
\end{align}
and 
\begin{align}
	\dot\Omega = - \frac{3 J_2}{2} n \left(\frac{R_\odot}{a}\right)^2 \left(\frac{\cos i}{(1 - e^2)^2}\right), 
	\label{eq:j2-node}
\end{align}
where $n = 2\pi/P$ is the mean motion of Sedna's orbit \citep{Brasser2006}. With $J_2 = 2.6149 \times 10^3$, the corresponding precession timescales are $t_{\rm apsidal} = 2\pi/\dot\omega \approx 1.4 \Gyr$,  $t_{\rm nodal} = 2\pi/\dot\Omega \approx 2.6 \Gyr$.
 
 The Kozai-Lidov timescale for Sedna due to the galactic potential is,
\begin{align}
	t_{\rm KL} \sim 4.0\left(\frac{0.1 M_\odot \textrm{pc}^{-3}}{\rho_G}\right) \left(\frac{10^4~\textrm{au}}{a}\right)^{\nicefrac{3}{2}}~\textrm{Gyr} = 360~\textrm{Gyr},
	\label{eq:KL-time}
\end{align}
where $\rho_G$ is the mass density of the Galaxy in the solar neighborhood, and the angular momentum diffusion timescale for Sedna from stellar passages is,
\begin{align}
	t_{\textrm{diff}} \sim 25\left(\frac{q}{30~\textrm{au}}\right) \left(\frac{10^4~\textrm{au}}{a}\right)^2~\textrm{Myr} = 25~\textrm{Gyr},
	\label{eq:diff-time}
\end{align}
with the timescales taken from \citet{Brasser2008}\footnote{The angular momentum diffusion timescale and Kozai-Lidov timescale were derived for $e\sim1$ orbits in \citet{Brasser2008}.}. 
The timescale for apsidal precession induced by the giant planets is the shortest of these three. The inclination e-folding timescale for Sedna is,
\begin{align}
	t_{\textrm{inst}} \sim 0.4~\frac{1} {2\pi} \left(\frac{M_\odot}{\Md}\right) P = 725 \left(\frac{M_\odot}{\Md}\right)~\textrm{yr}.
	\label{eq:ss-time}
\end{align}
If $\Md \approx 0.18 M_{\oplus}$, the instability timescale is of the same order of magnitude as the apsidal precession timescale.  
Thus, we find that in this scenario the self-gravity of small bodies in the Sedna region is the dominant dynamical driver\footnote{Of course this statement does not hold if a massive planet 9 \citep{Batygin2016} exists in this region.}. 

If we instead hypothesize that the minor planets in the Sedna region came from the scattered disk (a natural starting point for the instability, which requires only eccentric orbits in an axisymmetric disk), and detached themselves via self-gravitational torques, the more relevant comparison is that of the instability e-folding timescale to the energy diffusion time. 
The energy diffusion time due to planetary perturbations is given by \citet{Duncan1987}
 \begin{equation}
  t_{D(x)} \approx 1 \times 10^6 \left( \frac{10^4 \rm{au}}{a}\right)^{1/2} \left(\frac{10^{-4} \rm{au^{-1}}}{D(x)} \right)^2 \rm{yr},
\end{equation}
where $D(x)$ is the typical energy perturbation per perihelion passage. 
For a body originating in the scattered disk with perihelion distance of $p \approx 35 \rm{au}$ and a low orbital inclination, $D(x) \approx 10^{-5} \rm{au^{-1}}$ \citep[figure 1 of][]{Duncan1987}. This yields an energy diffusion time at Sedna's semi-major axis of $t_{D(x)} \approx 450 ~\rm{Myr}$, consistent with results from \citet{Bannister2017}. 
If $\Md \approx 0.5 M_{\oplus}$, the instability timescale equals the energy diffusion timescale.  
Thus in this scattered disk scenario, the self-gravity of small bodies in the Sedna region is the dominant dynamical driver if there exists about half an Earth mass of material at hundreds of au. 
In general, self-gravity in the inner Oort Cloud acts to increase the clearing timescale due to planetary perturbations. 
Gravitational torques increase the orbital inclination and the perihelion distance of minor planets, thus weakening the strength of scattering events. Minor planets can be more effectively retained.  

\citet{Fan2017} recently presented simulations of self-gravitating planetesimal disks within the context of the Nice Model. They found no evidence of the inclination instability, attributing its absence to mass depletion via ejections, and rapid precession of orbits due to the presence of the giant planets. 
We note however that the simulations ran for only $20 \Myr$ during the most violent phase of solar system evolution. We do not expect the inclination instability to dominate while the giant planets are migrating tens of au.
Furthermore, there is the important issue of particle number. If the original disk mass of $30 M_{\oplus}$ depletes to $\approx 0.5 M_{\oplus}$, the number of particles reduces from $N = 10^3$ to $N \approx 16$. The instability cannot be captured at this low particle number, as discussed in section~\ref{subsec:n-orbits}.
Thus, we cannot conclude from the \citet{Fan2017} results that the Nice model is incompatible with the inclination instability\footnote{It's interesting that the success ratio of the Nice model in reproducing the architecture of the current day solar system is a factor of $\sim$10 lower than in simulations without self-gravity.  Self-gravity of minor planets would appear to be a crucial component in the dynamical evolution of the solar system.}.

If the observed clustering of arguments of pericenter ($\omega$) of minor planets is a real effect \citep[see][]{Shankman2017b}, we need to either (i) be catching the instability during its linear phase, or (ii) be in the nonlinear, saturated phase and have something maintaining the clustering. It is interesting to note that the cone-shaped gravitational potential produced by the instability induces precession of eccentric orbits such that $\dot{\omega} < 0$, while the presence of the giant planets induces the opposite, $\dot{\omega} > 0$ (for $\cos i < 1/\sqrt{5}$; see equation~\ref{eq:j2-apse}). There may well be an interesting region of phase space where the two cancel and clustering is maintained (see \citet{Sefilian2018} for a similar idea involving longitude of perihelion ($\varpi$) clustering with a massive, apse-aligned disk). 
We have, so far, focused on the linear phase of the inclination instability in an idealized set-up. In future studies, we will directly apply the inclination instability to the solar system and examine its nonlinear evolution (Zderic et al. 2019, in preparation.). We will include the gravitational influence of the giant planets, and vary our initial conditions to reflect the scattered disk population and the inner Oort cloud formation in a stellar cluster. Finally, instead of presenting results averaged for many particles, we will focus on individual minor planets, particularly those that flip to retrograde orbits.

\acknowledgments
We thank the anonymous referee for their helpful comments. 
We gratefully acknowledge support from NASA Solar System Workings (SSW) under grant number 80NSSC17K0720.
Resources were provided by the NASA High-End Computing (HEC) Program through the NASA Advanced Supercomputing (NAS) Division at Ames Research Center. We also used the RMACC Summit supercomputer, which is supported by the National Science Foundation (awards ACI-1532235 and ACI-1532236), the University of Colorado Boulder, and Colorado State University. Simulations in this paper made use of the {\tt REBOUND} code which can be downloaded freely at https://github.com/hannorein/rebound. We used the open-source software TIOGA to make our plots.


\appendix
\label{conversions}
\section{Element Conversions}

The inclination instability is due
primarily to forces acting \textit{out} of the orbital plane.  These
forces exert torques on the orbit, ${\vec\tau} = {\vec r} \times {\vec
  f}= d{\vec j}/dt$, which result in rotations of the $\vec{j}$
vector.  Whereas none of the standard Kepler elements remain constant
during the instability, certain combinations of them do.  Hence in \citet{Madigan2016} we 
introduced two new coordinates that represent rotation
angles\footnote{The inclination angles \ia and \ib are defined in
  equation~\ref{eq:iaibie}; with this definition, \ia and
  \ib are independent quantities, but they are not formally rotation
  angles.  They are rotation angles to first order, however, so this
  difference does not affect our analysis in this paper.}  about the
semi-major axis of the orbit, \ia, and about the semi-minor
axis\footnote{Strictly speaking, an orbit does not rotate about its
  minor axis, rather about a chord parallel to its minor axis which
  runs through the focus of the ellipse, the latus rectum.} of the
orbit, \ib.  To complete a transformation from Kepler elements, we also
introduced \ie, the angle between the positive $\hat{x}$-axis and the
projection of the semi-major axis (or eccentricity vector) in the
$xy$-plane.  In the following conversions, we use $\hat{a}$ and $\hat{e}$ interchangeably as they are equivalent. \\

Inclination angles \ia, \ib, and \ie $\leftrightarrow$ orbit
vectors $\hat{a}$, $\hat{j}$, and $\hat{b}~(\equiv \hat{j} \times \hat{e})$:
\begin{subequations}
\begin{align}
  \cos i &= \cos \ia \cos \ib \left[1 - (\tan \ia \tan \ib)^2\right]^{1/2} \\
  \hat{a} &= \left[
    \begin{array}{c}
      \cos \ib \cos \ie \\
      \cos \ib \sin \ie \\
      -\sin \ib
  \end{array}
  \right]\\
  \hat{b} &= \left[
    \begin{array}{c}
      \tan \ib \sin \ia \cos \ie - ({\cos i}/{\cos \ib}) \sin \ie \\
      \tan \ib \sin \ia \sin \ie + ({\cos i}/{\cos \ib}) \cos \ie \\
      \sin \ia
    \end{array}
    \right] \\
  \hat{j} &= \left[
    \begin{array}{c}
      \tan \ib \cos i \cos \ie + ({\sin \ia}/{\cos \ib}) \sin \ie \\
      \tan \ib \cos i \sin \ie - ({\sin \ia}/{\cos \ib}) \cos \ie \\
      \cos i
    \end{array}
    \right]
\end{align}
\end{subequations}

Kepler elements $\leftrightarrow$ orbit vectors $\hat{e}$, $\hat{j}$, and $\hat{b}$:
\begin{subequations}
\begin{align}
  \hat{\vec{e}} &= 
  \left(
  \begin{array}{c}
    \cos \omega \cos \Omega - \sin \omega \cos i \sin \Omega \\
    \cos \omega \sin \Omega + \sin \omega \cos i \cos \Omega \\
    \sin \omega \sin i
  \end{array}
  \right) \\
  \hat{\vec{j}} &= 
  \left(
  \begin{array}{c}
     \sin i \sin \Omega \\
    -\sin i \cos \Omega \\
     \cos i
  \end{array}
  \right)\\
  \hat{\vec{b}} &= 
  \left(
  \begin{array}{c}
    -\sin \omega \cos \Omega - \cos \omega \cos i \sin \Omega \\
    -\sin \omega \sin \Omega + \cos \omega \cos i \cos \Omega \\
     \sin i \cos \omega
  \end{array}
  \right)
\end{align}
\end{subequations}

\begin{subequations}
\begin{align}
  \vec{n} &= \hat{z}\times\vec{j} \\
  i &= \arccos\left[\oldhat{j}_{\text{z}}/j\right] \\
  \Omega &= \arctan\left[\oldhat{n}_{\text{y}}, 
                         \oldhat{n}_{\text{x}}\right] \\
  \omega &= \arctan\left[\text{sgn}(\oldhat{e}_{\text{z}}) 
                         \left[1 - (\hat{n}\cdot\hat{e})^2\right]^{1/2},
                         \hat{n}\cdot\hat{e}\right]
\end{align}
\end{subequations}

Kepler elements $\leftrightarrow$ inclination angles \ia, \ib, and \ie (note that these definitions are valid for prograde orbits only):

\begin{subequations}
\begin{align}
  \ia &= \arcsin(\sin i \cos \omega) \\
  \ib &= \arcsin(-\sin i \sin \omega) \\
  \ie &= \Omega - \arctan\left[\cos \omega, \cos i \sin \omega\right]
  + \pi/2
\end{align}
\end{subequations}

\begin{subequations}
\begin{align}
  i &= \arccos\left[\cos \ia \cos \ib \left[1 - (\tan \ia \tan
      \ib)^2\right]^{1/2}\right] \\
  \omega &= \arctan\left[-\text{sgn}(\ib) \sqrt{1-\left(\frac{\sin \ia}{\sin i}\right)^2},
    \frac{\sin \ia}{\sin i}\right] \\
  \Omega &= \ie + \arctan\left[\tan \ib \cos i, \frac{\sin \ia}{\cos \ib}\right]
\end{align}
\end{subequations}

\end{document}